# Kinematics and Formation Mechanisms of High-Redshift Galaxies

David R. Law (UC Los Angeles; drlaw@astro.ucla.edu),
Shelley A. Wright (UC Irvine), Richard S. Ellis (Caltech), Dawn K. Erb (UC Santa Barbara),
Nicole Nesvadba (Observatoire de Paris), Charles C. Steidel (Caltech),
Mark Swinbank (Cambridge)

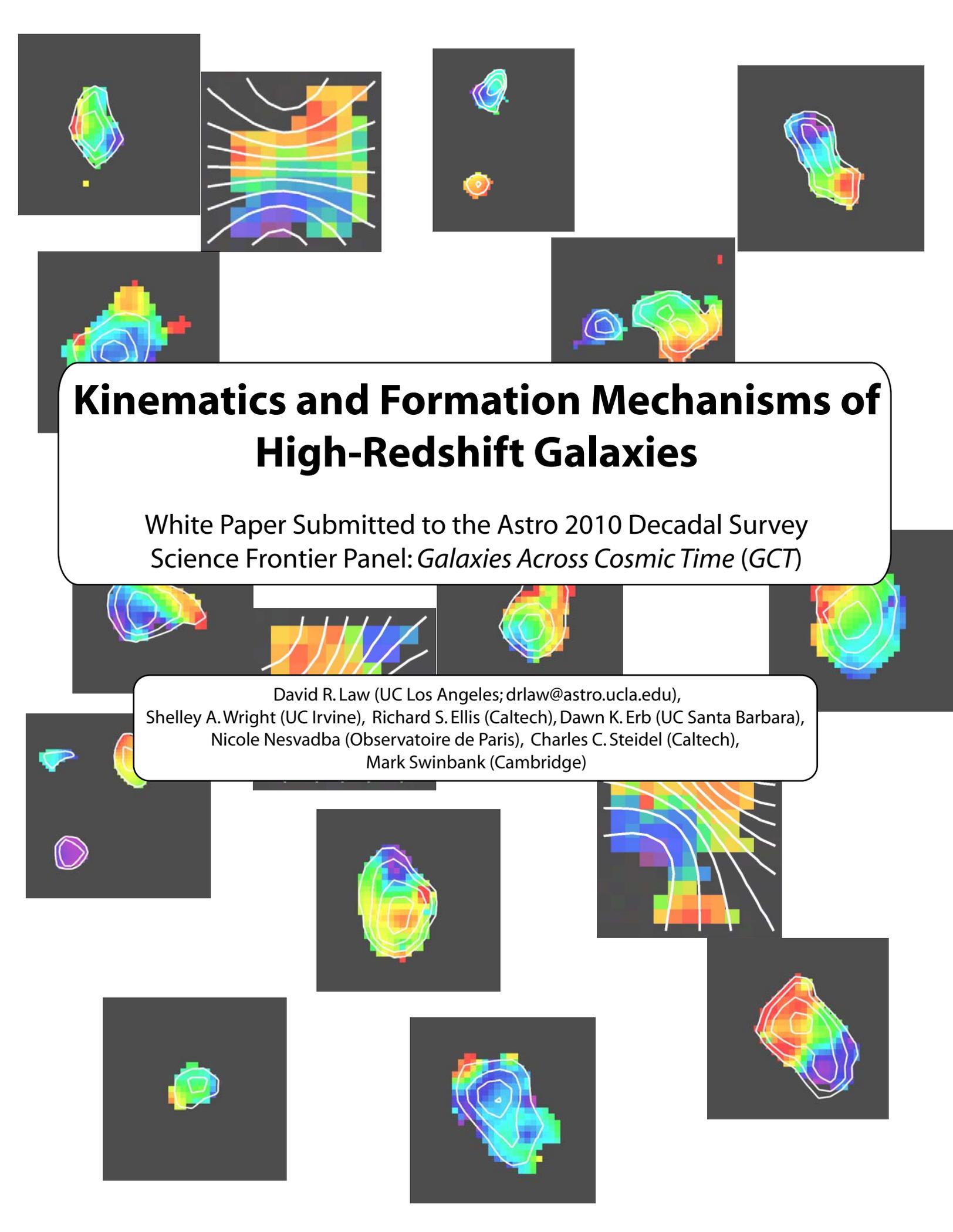

# 1. Background:

In recent years our understanding of the broad global characteristics of galaxies in the young universe has grown considerably. Using rest-frame UV and optical spectroscopy and multi-wavelength broadband photometry it has been possible to estimate their stellar and dynamical masses, metallicities, ages, and star formation rates at redshifts $z < 1.5$ (Cowie et al. 2008; and references therein), $z \sim 1.5 - 3$ (Shapley et al. 2005; Erb et al. 2006; Reddy et al. 2006; Papovich et al. 2006), and $z > 4$ (e.g., Stark et al. 2007; Verma et al. 2007). Such studies indicate that the majority of structures observed in the local universe were already in place at $z \sim 1$ (Papovich et al. 2005) and point to the era spanned by the redshift range $1.5 \leq z \leq 3$ as the peak epoch of both star formation (Dickinson et al. 2003) and AGN activity in the universe.

Despite our knowledge of the global characteristics of galaxies at such redshifts however, our understanding of their internal structure and dynamical evolution has been limited by their small angular size, typically $\leq 1$ arcsecond. Such objects are generally not well resolved by the ground-based imaging and spectroscopy which form the backbone of the observational data. This limitation precludes us from addressing questions such as: (i) What are the triggering and regulation mechanisms of their starbursts? (ii) Is the gas that fuels these starbursts accreted in discrete gas-rich major mergers, by gradual adiabatic accretion from an extended gaseous halo, or directly from cosmological filaments? Each of these distinctions has important implications for the evolution and development of structure and stellar populations within galaxies.

In the local universe, the morphology of a galaxy is often one of the most accessible observables, and is closely correlated with a wealth of information about a galaxy's kinematics, rate of star formation, and recent merger history. One popular approach to studying high-redshift galaxies has therefore been to use high-resolution imaging from space-based telescopes to quantify the morphological characteristics of these galaxies, and such studies have often concluded that the merger fraction was much higher at these redshifts than today (e.g., Lotz et al. 2006; Conselice et al. 2007; although c.f. Conroy et al. 2008). However, the high-resolution imaging of $z \sim 2 - 4$ galaxies reveals a population of extremely irregular, clumpy galaxies that bear little similarity to the local Hubble sequence (Figure 1; see also Giavalisco et al. 1996; Lotz et al. 2006), and some recent studies suggest that there may be few unambiguous relations between morphology and kinematics, star formation properties, etc., for these galaxies (Law et al. 2007b; Vanzella et al. 2009). Indeed, for a galaxy with multiple clumps in the light distribution it is typically not possible to distinguish between star formation in a clumpy gaseous disk or a high-speed merger between two discrete (albeit close in projected distance) galaxies from morphological data alone.

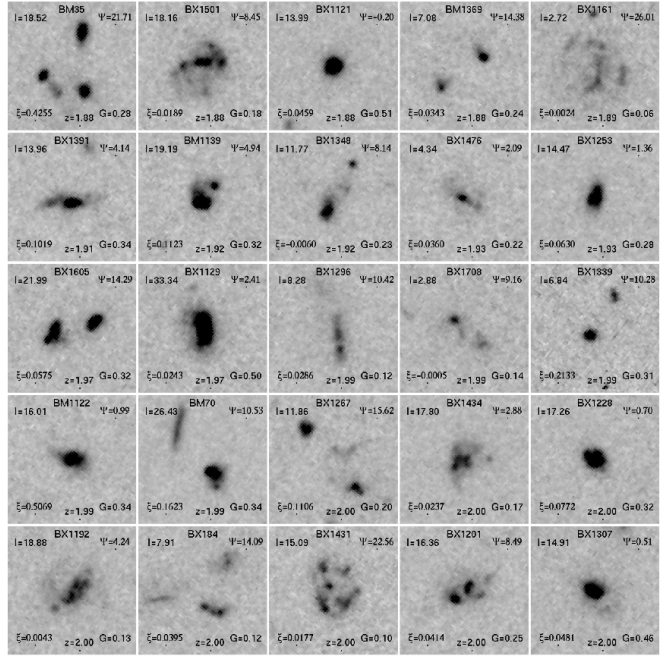

Figure 1: *HST/ACS images of 25 galaxies at z ~ 2 in the GOODS-N field, each box is 3 arcsec on a side (from Law et al. 2007b). Note the complex morphologies of these systems.*

## 2. Kinematics:

Given the complications of such morphological studies, invaluable additional information can be gleaned from the nebular emission lines (e.g., Hα, Hβ, [O III], [O II], and [N II]) which provide good kinematic tracers of the ionized gas surrounding active star forming regions. Such nebular emission-line spectroscopy has been used (e.g.) to trace the evolution in the Tully-Fisher relation out to z ~ 1.2 (Weiner et al. 2006; Kassin et al. 2007) and suggests the growing importance of non-circular motions to this relation with increasing redshift. In combination with new adaptive optics (AO; e.g. Wizinowich et al. 2006) and integral-field unit (IFU) technologies it has additionally become possible for ground-based telescopes to overcome the limitations imposed by atmospheric turbulence and "dissect" galaxies with spectroscopy on hitherto-unprobed sub-kiloparsec scales.

IFU studies at z ~ 1.5 (Wright et al. 2007, 2009) confirm a slight increase in non-circular motions relative to the local universe, but also find evidence of organized rotation within their galaxy sample. At higher redshifts, numerous studies at z ~ 2 - 3 (e.g., Förster-Schreiber et al. 2006; Genzel et al. 2008; Law et al. 2007a, 2009; Nesvadba et al. 2008) have found that galaxies have extremely large velocity dispersions ($\sigma \sim 80$ km s$^{-1}$) as compared to their rotational velocity (V) about a preferred kinematic axis. While exact values of the ratio V/σ vary from less than unity up to about 4 – 5, there is clearly a dynamical difference from disk galaxies in the nearby universe, which typically have V/σ ~ 15 – 20 (e.g., Dib et al. 2006).

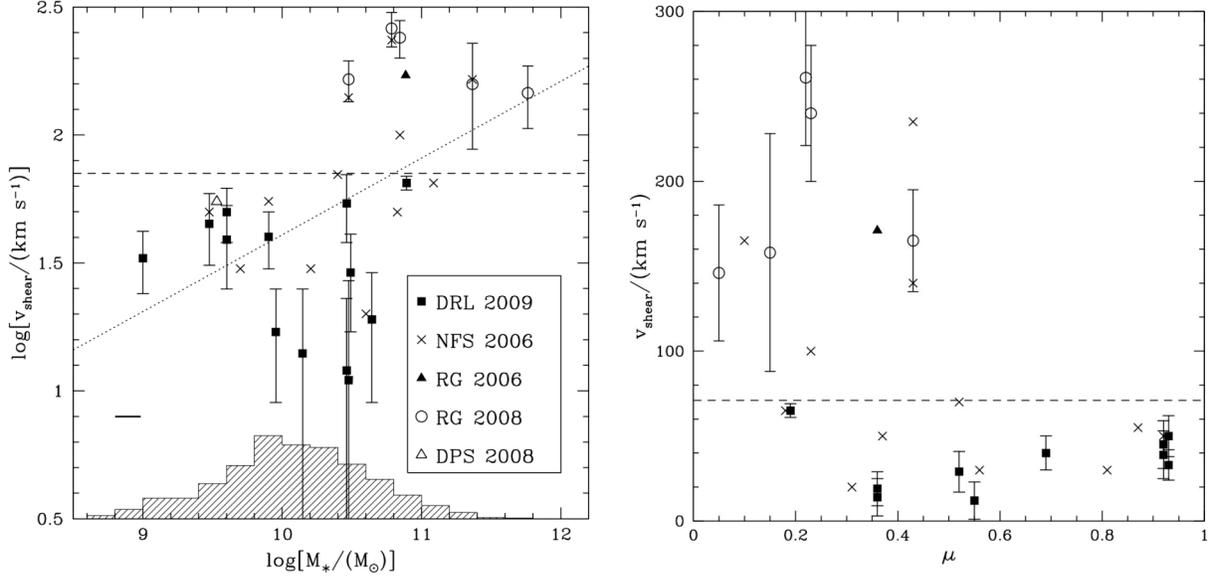

Figure 2: *Plot of stellar mass ($M_*$) and gas fraction (μ) vs. maximum line-of-sight velocity shear for a selection of recent observational samples (from Law et al. 2009). The dashed line indicates the mean velocity dispersion of the galaxies, the dotted line is a least-squares fit to the data. The histogram represents the relative number of galaxies in each logarithmic mass bin for a sample of 818 galaxies in the redshift range z = 1.8 - 2.6. Uncertainties in $v_{shear}$ are given where available, the solid line in the lower left corner of the plot indicates the typical uncertainty in $M_*$.*

Such evidence for turbulent star formation in these young galaxies led Genzel et al. (2008) to propose that the buildup of central disks and bulges at early times may be driven by secular evolution in gas-rich disks. Additional recent evidence (Law et al. 2009) also suggests that for the majority of star-forming galaxies at z ~ 2 the kinematics are not merely influenced by velocity dispersions but are instead dynamically dominated by them. As illustrated in Figure 2, $V/\sigma < 1$ (i.e. similar to bulge-like values) for galaxies of typical stellar mass ~ $10^{10}\,M_\odot$ which tend to be gravitationally dominated by their cold gas.

Such dispersion-dominated kinematics are at odds with the classic picture of galaxy formation (e.g., Mo, Mao, & White 1998), according to which we might expect to see rapid star formation occuring in rotationally-supported disks composed of gas accreted relatively adiabatically from the surrounding halo. Instead, the lack of rotation observed in typical galaxies and the turbulent nature of the disks which may be more common at higher stellar masses suggests an alternate formation mechanism. Rather than forming their early stellar populations in rotationally supported disks, these young galaxies may be forming stars in violently relaxing low-angular momentum gas originating from instabilities in these extremely gas-rich systems (e.g. Bournaud et al. 2007) or direct accretion from cosmological filaments via cold flows (e.g. Dekel & Birnboim 2006, Kereš et al. 2005). The detailed physical structure which might result from such models is unclear, although some recent theoretical work suggests (Dekel et al. 2009) that clumpy accretion may naturally give rise to a highly turbulent system which might form compact stellar spheroids at relatively high redshift.

Gas accretion via such cold-mode mechanisms has substantial implications for the development of structure as a function of both redshift and environment as the relative importance of cold flows vs. hot accretion is a function of both galaxy and halo mass (e.g., Kereš et al. 2005). Similarly, revisions to the hot gas infall rate may change constraints on AGN feedback models and the balance of Ly$\alpha$ versus X-ray cooling mechanisms (e.g., Fardal et al. 2001). Such a confluence of theoretical revisions may in turn suffice to explain the observed changes in the cosmic star formation history, providing a natural framework in which to understand the fundamental differences between local Hubble-type galaxies and the irregular starbursts that formed the majority of the stellar matter in the universe.

What is immediately clear, however, is that we do not yet understand the dynamical state of galaxies during the period when they are forming the bulk of their stars. High gas-phase velocity dispersions (and small velocity gradients) appear to be a natural consequence of the instabilities resulting when cold gas becomes dynamically dominant, as it seems to be in the central few kiloparsecs of a large fraction of star-forming galaxies at z ~ 2 - 3. Learning more precisely how and why this occurs may be the key to understanding galaxy formation.

## 3. Future Development:

The work described above has demonstrated the real and present capability of observationally constraining theoretical models of galaxy formation at the epoch when the universe formed the bulk of its stars, highlighting the need for a theoretical understanding of cold gas accretion and stability on sub-kiloparsec scales. Additionally, we are now able to frame specific observational questions regarding the nature of galaxy formation in the young universe. In particular,

    1) Are the high observed random velocity components primarily a selection effect of observing the most rapidly star-forming (and therefore perhaps most turbulent) galaxies at a particular epoch? As illustrated in Figure 3, current observations probe the high surface-brightness end of the galaxy population. Might the ionized gas kinematics of these galaxies differ from lower surface brightness galaxies at similar redshifts?

    2) Observationally, the flux from many galaxies appears dominated by the small V/$\sigma$ of ionized gas within the central 2 - 4 kpc of the galaxy. Is there also low surface-brightness star formation at larger radii, and (if so) is the gas at these radii rotationally supported or does it have a similarly small V/$\sigma$ (potentially discriminating between cold accretion and disk instabilities)?

    3) In a select few cases, observations of gravitationally lensed galaxies (e.g. Nesvadba et al. 2006, Stark et al. 2008) have demonstrated that some galaxies have appreciable kinematic structure on ultra-fine scales (~ 100 pc) unresolvable even with current AO technologies. Using a combination of improved spatial resolution and greater light-gathering capability, it will be possible to obtain a more detailed picture of the underlying dynamical mechanisms driving the observed dispersion-dominated kinematics, possibly resolving stellar populations into discrete super-star clusters akin to those observed in local starburst galaxies (e.g., Overzier et al. 2008) and determining the role of starburst feedback within H II regions.

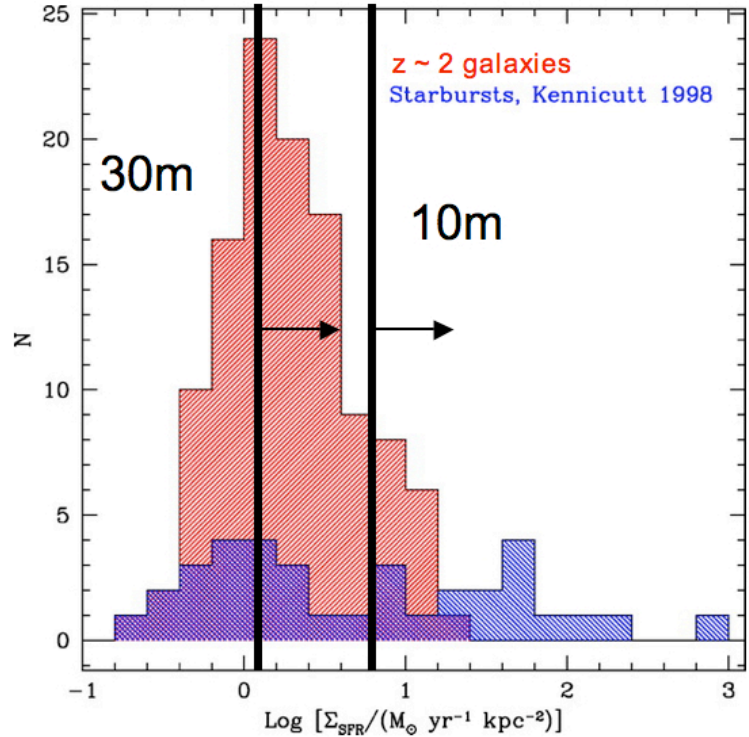

Figure 3: *Distribution of star-formation rate densities for a z ~ 2 rest-UV selected galaxy sample compared to local starbursts (from Erb et al. 2006). Overplotted are black lines indicating the typical star-formation rate density probed by current AO/IFU surveys with a 10m primary mirror and those pertaining to future 30m-class facilities.*

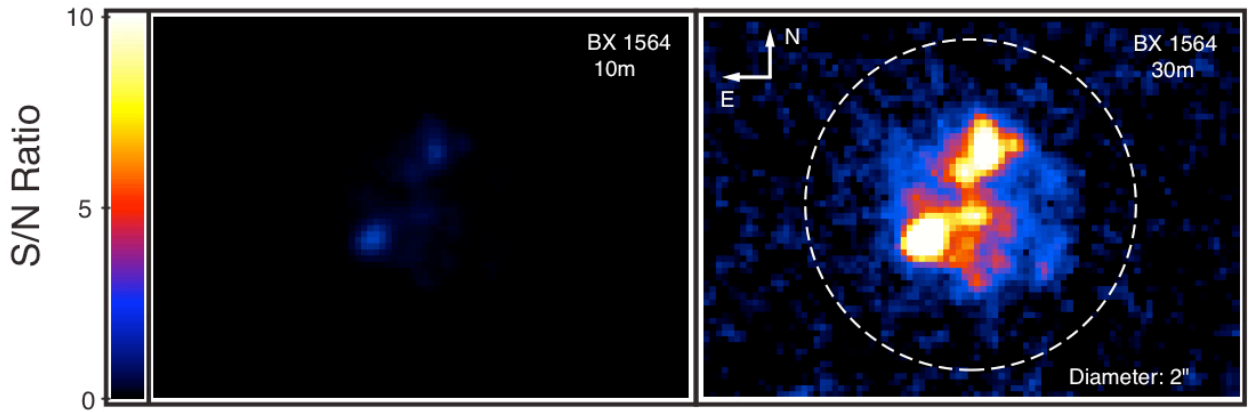

Figure 4: *Relative signal-to-noise ratios expected for a two-hour observation of Hα emission from the star-forming galaxy HDF-BX1564 with current 10m (left-hand panel) and next-generation 30m (right-hand panel) facilities. Note the key increase in ability to detect low surface-brightness emission on small angular scales between the two primary components.*

4) What is the relation between typical star forming galaxies and faint AGN/QSOs? Can active star-forming regions be seen in the vicinity of these active nuclei, and (if so) how do their properties differ from those of the rest of the galaxy population? Spatially resolved maps of diagnostic line ratios involving weaker transitions (such as [O II] ] λ 3726/3729, [O I] λ 6300, and [S II] λ 6716/6731) will be helpful in identifying and characterizing weakly active nuclei.

5) Does the decrease in V/σ with increasing redshift continue above z ~ 3? While the strongest nebular emission lines are redshifted to an inaccessible region of the near-IR at z > 3.4 shorter wavelength transitions such as [O II] λ 3726/3729 can be observed to z ~ 5.5 (e.g., Swinbank et al. 2007).

6) Star formation within galaxies is fueled by reservoirs of cold molecular hydrogen, which is typically assumed to correlate with observed star formation via an extension of the local Schmidt-Kennicutt law (Kennicutt 1998). Using ultra-sensitive millimeter-wavelength facilities such as the Atacama Large Millimeter Array (ALMA) it will be possible to verify whether young galaxies at z ≥ 2 experience modes of star formation consistent with local spiral galaxies (where the star formation occurs within giant molecular clouds; e.g., Daddi et al. 2008) or are so perturbed (as suggested by V/σ) that the star formation efficiency is much higher, akin to local LIRGS and ULIRGS.

As demonstrated by numerical simulations shown in Figure 4, the combination of diffraction-limited angular resolution (~ 15 milliarcseconds at λ = 2 μm, corresponding to ~ 100 pc at z = 2), greater light-gathering capacity, and potentially greater multiplexing capabilities of future 30m-class near-IR telescopes will permit mapping of nebular line emission from galaxies in the young universe on scales presently impossible without the aid of gravitational lensing. Additionally, the synergistic combination of high angular resolution maps of rest-frame optical nebular emission (tracing extended warm ionized gas) and millimeter-wavelength CO emission (tracing compact cold molecular gas) will make possible a deeper, more complete picture of the role of gas in early galaxies and its connection with star formation. Such an investigation of the relation between small-scale processes (such as star formation rate densities, starburst feedback, and evolution in the characteristic sizes of individual star forming regions) and the resulting global galactic properties will be of critical importance in developing a broad and consistent picture of the relevant baryonic physics during the galaxy formation process in the young universe.